# Quantum Flexoelectricity in Low Dimensional Systems

Sergei V. Kalinin[*] and Vincent Meunier[†]

Oak Ridge National Laboratory, Oak Ridge, TN 37831

## Abstract

Symmetry breaking at surfaces and interfaces and the capability to support large strain gradients in nanoscale systems enable new forms of electromechanical coupling. Here we introduce the concept of *quantum flexoelectricity*, a phenomenon that is manifested when the mechanical deformation of non-polar quantum systems results in the emergence of net dipole moments and hence linear electromechanical coupling proportional to local curvature. The concept is illustrated in carbon systems, including polyacetylene and nano graphitic ribbons. Using density functional theory calculations for systems made of up to 400 atoms, we determine the flexoelectric coefficients to be of the order of ~ 0.1 e, in agreement with the prediction of linear theory. The implications of quantum flexoelectricity on electromechanical device applications, and physics of carbon based materials are discussed.

[*] sergei2@ornl.gov
[†] meunierv@ornl.gov

Since the seminal work of Iijima,[1] carbon nanotubes have emerged as prototype 1D systems and ideal test-beds for studies of quantum confinement effects through the macroscopically accessible transport properties.[2] Multiple studies on ballistic conductance,[3] simple and double quantum dot states,[4] Kondo effect and spin injection,[5,6] Luttinger liquid behavior,[7] and mechanical control of electronic structure,[8] have been reported. In recent years, this interest has extended to the 2D crystals, including graphene and layered oxides. The 2D nature of graphene gives rise not only to new physics but also to novel device applications, including spintronic devices,[9] electron lenses[10] and electromechanical switches.[11] Here, we predict the unusual electromechanical coupling effects that arise in graphene and are forbidden by rotational symmetry of carbon nanotubes.

In the last two decades, the fact that "small is different" has been established for a wide variety of phenomena, including electrical, optical, magnetic, and mechanical properties of materials. The emergence of unusual phenomena has given rise to prospective areas of research in the physical and biological sciences such as nanomechanics,[12] plasmonics,[13] nano-optics,[14] and molecular electronics,[15] as well as to multiple new material and device applications. However, one largely untapped but potentially very important area of nanoscience involves the interplay of electricity and mechanics at the nanoscale. Linear piezoelectric coupling between electric field and stress is extremely common in inorganic materials (20 out of 32 symmetry groups are piezoelectric)[16] and is nearly universal in biopolymers.[17] The origins for piezoelectricity can be traced to a single chemical bond level[18] and the properties of bulk materials are defined by relative bond orientations and collective interactions. In particular, in centrosymmetric materials the presence of the center of inversion results in the absence of bulk piezoelectric properties. However, the symmetry breaking on surfaces and interfaces in non-polar materials can allow new forms of electromechanical coupling forbidden in the bulk, such as surface piezoelectricity and flexoelectricity.[19] The constitutive relations for direct flexoelectric effect is

$$P_s = f_D(c_1 + c_2),\tag{1}$$

where $c_1$ and $c_2$ are the principal (Gaussian) curvatures, $P_s$ is the polarization [C/m], and $f$ is the flexoelectric constant [C]. The inverse flexoeffect is

$$c_1 + c_2 = (f_C/D)E,\tag{2}$$

where $D$ is the bending stiffness and $E$ is the electric field. From thermodynamic Maxwell relations, the flexoelectric coefficients in Eqs. (1) and (2) are equal, $f_C = f_D = f$. In comparison, in piezoelectricity, the application of stress, $X$, results in the electrical polarization (direct piezoelectric effect, $P = dX$), while application of electric field, $E$, results in strain (converse piezoelectric effect, $x = dE$), where $d$ is the piezolectric constant.

The flexoelectric constant is related to the charge distribution within the system as

$$f = \int \left.\frac{\partial P(z,c)}{\partial c}\right|_{c=0} dz,\tag{3}$$

where $P(z,c)$ is the curvature-dependent polarization distribution in the 2D object and $z$ is the normal coordinate.

While flexoelectricity is relatively unusual in inorganic materials, it is extremely common in soft condensed matter systems such as liquid crystals and cellular membranes. In particular, membrane flexoelectricity[20] forms the functional basis for processes such as acoustic wave detection in outer hair cell stereocilia (hearing)[21] and energy storage in mitochondria. The mechanism leading to flexoelectricity in membranes and liquid crystals can be interpreted as follows. In the flat state, there is no dipole moment across the membrane.

Upon bending, redistribution of ions and charges results in the formation of a net dipole moment across the membrane, and hence of a coupling with electric field that increases with curvature. Notably, similar behavior can be anticipated in 2D and 1D quantum systems such as graphene, in which the increase of curvature causes a redistribution of the electron gas in the normal direction, resulting in local electric dipole and hence flexoelectric coupling. We refer to this phenomenon as *quantum flexoelectricity*.

In this letter, we utilize first principles density functional theory (DFT) calculations and analytical models to study quantum flexoelectricity in model graphene and polyacetylene systems. All-electron DFT calculations within local density approximation of various forms of flexed graphene membranes and polyacetylene were performed using the quantum chemistry package NWchem[22] in the 3-21G (large 2D systems)[23] and 6-31G* (1D systems)[24] atom-centered, contracted Gaussian basis sets. The convergence of 3-21G results were tested for a number of configurations using the more complete 6-31G* basis. For large systems, such as the ones studied here, it is not computationally feasible to systematically include more functions, since it would lead to catastrophic numerical linear dependence. However, in contrast to small molecules, the response of the systems studied here to the uniform applied electric field is neither primarily on-site nor due to the long tail of the charge density, but arises from the redistribution of charge between sites. The electric field was introduced in the calculation by placing point charges (with opposite charge) far away from the systems, in such a way as to impose a uniform and constant field over the whole region around the structure. The dipole moment $\mu$ was obtained by fitting the energy curve $E(\varepsilon)$ as function of electric field $(0,0,\varepsilon_z)$ by

$$E = E_0 + \mu\varepsilon_z + P_2\varepsilon_z^2 + O(\varepsilon_z^3) \tag{1}$$

As an example of a 1D system, we have chosen a polyacetylene molecule, formed by 20 carbon atoms and 4 terminating hydrogen atoms. The molecule was progressively curved, from linear (infinite radius of curvature) to almost circular arrangement. In each step, the C-C and C-H bond length were kept at equilibrium distances of 0.14 and 0.11 nm, respectively. Shown in Fig. 1 is the schematics of the linear molecule in the uniform field and the dependence of the total energy of the system on the radius of curvature for different fields. For large radii of curvature, the free energy decreases uniformly with electric field due to the monotonic increase of bending-induced dipole moment. As expected, the dipole moment converges to 0.0 for very large radius of curvature. Interestingly, for small radii of curvature the free energy and dipole moment dependence becomes non-monotonic, exhibiting a clear minimum at around 0.8 nm. This behavior originates from the fact that for radius of curvature below which the hydrogenated ends of the molecule are diametrically opposed, negative contributions reduce the overall dipole moment perpendicular to the chain, eventually leading to zero dipole when the two ends meet (not shown).

The immediate consequence of data in Fig. 1 is that in the strong static electric fields the equilibrium state of the molecule corresponds to the bent configuration. At that position, the repulsive energy terms associated to the end-end interaction and the elastic stress are compensated by the attractive dipole-electric energy due to charge redistribution. It is interesting to note that when immersed in an external electric field, the stable position corresponds to a curved atomic chain. Therefore, the system can be electromagnetically excited into the bent configuration and, after switching off the external electric source, a clear signal corresponding to the collective relaxations of the molecules should be observable, unless conformational changes lead to local minima during relaxation. Such a measurement would yield information on intrinsic properties of the un-excited system.

As an example of a 2D system, a single graphene sheet ($sp^2$ hexagonal network with edge carbon atoms saturated with hydrogen) was used. The system was curved onto a cylinder, along zig-zag and armchair edges, respectively. In each case, 4 different radii of curvature (0.81, 1.62, 3.24, 6.48 nm for the armchair systems and 0.78, 1.57, 3.13, and 6.26 nm for the zigzag systems) were considered. The resulting systems are equivalent to finite carbon nanotubes open along their axis, with H-saturated edges; or alternatively to finite nano graphitic ribbons (NGR). In both cases, the deformation proceeds well below the plastic regime (no formation of 5 or 7 membered rings), therefore keeping the three-fold coordination of each carbon atoms. After convergence down to $10^{-6}$ a.u of the total energy, the dipole moment $\mu$ was obtained from the multipole analysis of the converged electronic density. Evolution of the graphene membranes in computational experiments similar to the one performed for the linear carbon chain is shown in Fig. 2 (a). Here, the atomic positions of the atoms are adjusted in such a way that they fall on the cylinder with radius $R$. The evolution of the dipole moment with radius of curvature is shown in Figs. 2 (b,d). The calculated dependence is

$$p = a/R \qquad (2)$$

for large radii of curvature and weakly depends on the type of graphene sheet. The behavior deviates for smallest radius, due to the interaction between the two ends of the sheet. For smaller radii, the sheet closes more and more and eventually will close on a seamless cylinder (not shown), at which point the dipole moment will vanish. Therefore, the $p$ vs. $R$ relationship is bounded by $p = 0$ for both small and large radii. In the regions of small curvatures, i.e.

where Eq. (2) is applicable, the fit of DFT data leads to coefficient $a = 0.71 \pm 0.1$ e nm$^2$ for zig-zag and $a = 0.51 \pm 0.1$ e nm$^2$ for armchair geometries. Notably, this behavior is similar to the one predicted for membrane flexoelectricity, $P_s = f(c_1 + c_2)$. Here the principal Gaussian curvatures are $c_1=1/R$ and $c_2=0$. It follows that flexoelectric coefficient is $f = 0.00187 - 0.00157$ e nm$^2$/atom depending on the system.

To rationalize these observations, we consider simple linear theory for flexoelectric coupling in quantum systems. The $p_z$-electron system of the graphene sheet is represented as two sheets of the two-dimensional electron gas (2DEG) located at distance, $t$, from the plane of carbon atoms. Here, $t=h/2$ can be defined as half the thickness $h$ of the slab representing the electron distribution around the graphene sheet. An example of charge distribution is shown on Fig. 3. Upon bending the change in area of the outer shell is $S^+ = S^0(1+t/R_0)$ and for internal shell is $S^- = S^0(1-t/R_0)$, where $S^+$, $S^-$, and $S^0$, are the area of surface element on the outer and inner 2DEG sheets and graphene membrane *per se*, and $R_0$ is the radius of curvature. Notice that the overall area does not change upon bending, $S^+ + S^- = 2S^0$. Hence, the electron redistribution between top and bottom surfaces is local in this approximation and does not result in charge redistribution across the conjugate system, which will eventually have to be precluded by long-range electrostatic interactions. Note that this approximation is well justified after inspection of charge redistribution calculated by DFT (Fig. 3). It follows tat the transferred charge across the membrane per atom is simply given by the change in surface area: $\delta q = (1/2 e^-) t/R_0$ and the induced dipole moment is $p = 2t\delta q = e^- t^2/R_0$. Estimating the mean graphene thickness at about 0.08 nm,[25] the flexoelectric coefficient $f$ is found to be about 0.0016 e nm$^2$/atom. Finally, the dipole moment density, i.e. polarization, can be estimated as $P_s = e^- t^2 / (a_{c-c}^2 \sqrt{3}/2)(1/R_0)$, where $a_{c-c} = 0.14$ nm is the carbon-carbon bond length. Hence, the corresponding flexoelectric coefficient is $f_0 = 0.094$ e$^-$ (theory) and 0.11 e$^-$ (DFT). Note that DFT and simplified theory agree strikingly well. In comparison, typical flexoelectric coefficients in biological membranes is 0.1 – 10 e$^-$ depending on mechanism.[20] Flexoelectric coefficient determined from the DFT is close to the anticipated one in bulk materials; however, the ability of low-dimensional systems to sustain much larger strain gradients compared to bulk materials renders this effect much more relevant on the nanoscale.

To summarize, in this letter we have predicted a general form of electromechanical coupling in low-dimensional systems – quantum flexoelectricity. The redistribution of the 2D electron gas density in the normal direction due to curvature results in local dipoles, and hence electromechanical response. We anticipate that this behavior is universal for all low-dimensional systems that can support large deformations. As applied for carbon-based nanostructures, flexoelectric coupling can be used as a functional basis for novel electric field and mechanical sensors and actuators and nanoelectromechanical systems. Furthermore, flexoelectric coupling should result in strong electromechanical couplings in the disordered systems based on graphene (carbon foams), since there presence of curved regions will result in strong built-in strain gradients. The quadratic dependence of flexoelectric coupling on the effective thickness favors strong flexoelectric coupling in e.g. free-standing metal films, etc.

Quantum flexoelectricity will significantly affect shape fluctuation of 2D materials in perpendicular electric field, favoring fluctuations parallel to the field and suppressing antiparallel contributions, giving rise to macroscopic deformation even on nominally flat surface. This behavior can enable new classes of macroscopic electromechanically active materials and nanodevices, e.g. based on grapheme and carbon nanofoams.

Research supported by Oak Ridge National Laboratory, managed by UT-Battelle, LLC, for the U.S. Department of Energy under Contract DE-AC05-00OR22725. A Part of this work was supported by the Center for Nanophase Materials Sciences (CNMS), sponsored by the division of Scientific User Facility, U.S. Department of Energy.

**Figure Captions**

**Fig. 1.** (a) Schematics of polyacetylene chain for 6 different radii of curvature. (b) Variation of net dipole moment of the molecule as a function of radius of curvature. (c) Dependence of free energy (solid) for different electric fields as a function of radius of curvature. The amplitude of the applied electric field is given on the figure.

**Fig. 2.** (a)-(c) Schematics of flexed graphitic membranes with armchair and zigzag edges along the circumference. (b)-(d) Corresponding evolution of dipole moment dependence as a function of the radius of curvature R. Fit of the curves f ~ 1/R is shown by the solid line.

**Fig. 3.** Top cartoon represents the density distribution on an horizontal plane passing through top atoms of an armchair terminated rolled graphitic membrane at a radius of curvature of 6.5 nm. Bottom: isosurface plot of the electronic charge density for the same system.

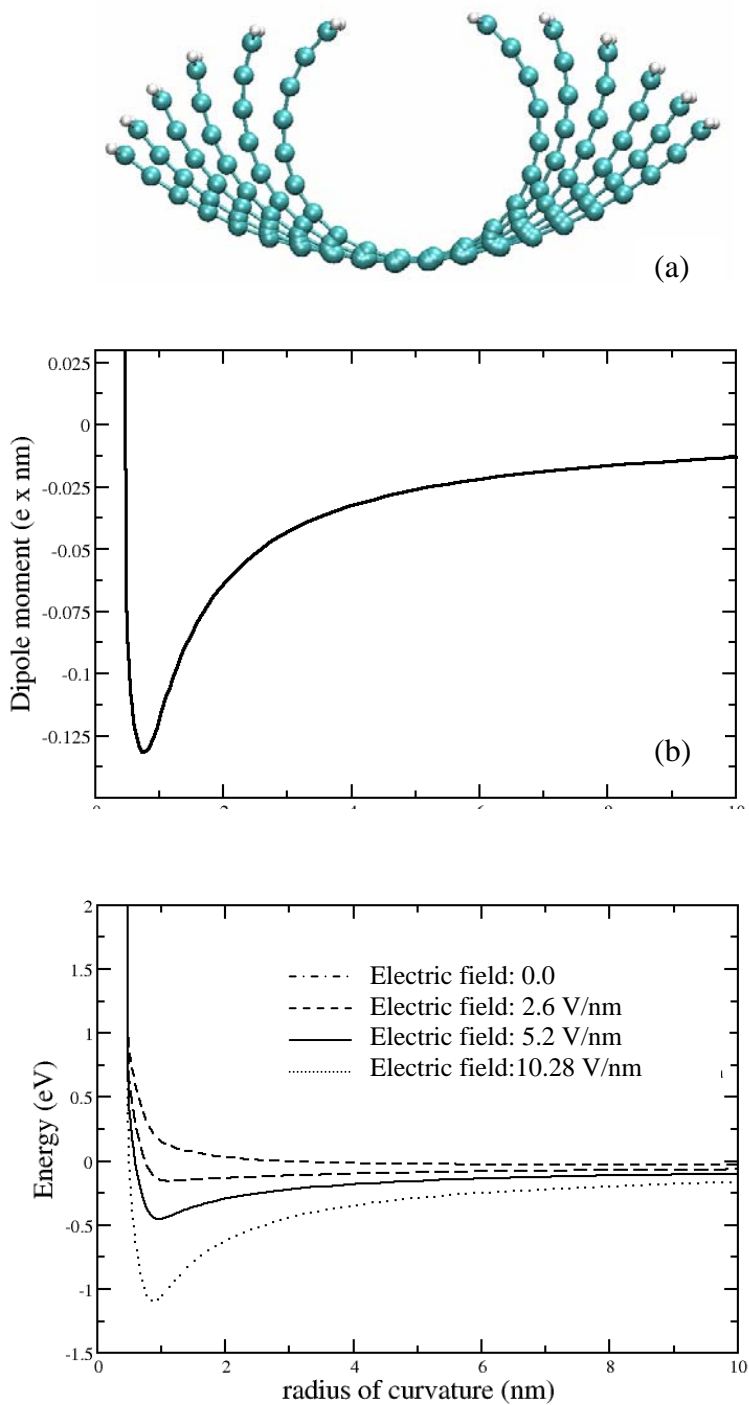

**Figure 1.** S.V. Kalinin and V. Meunier

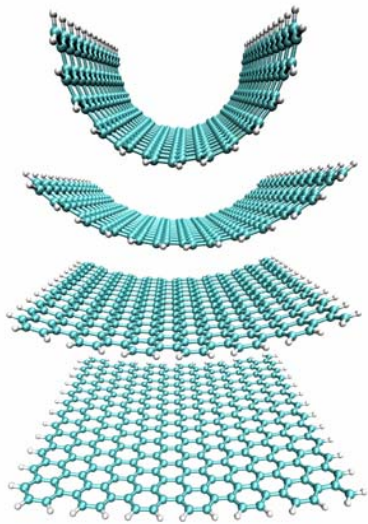
(a)
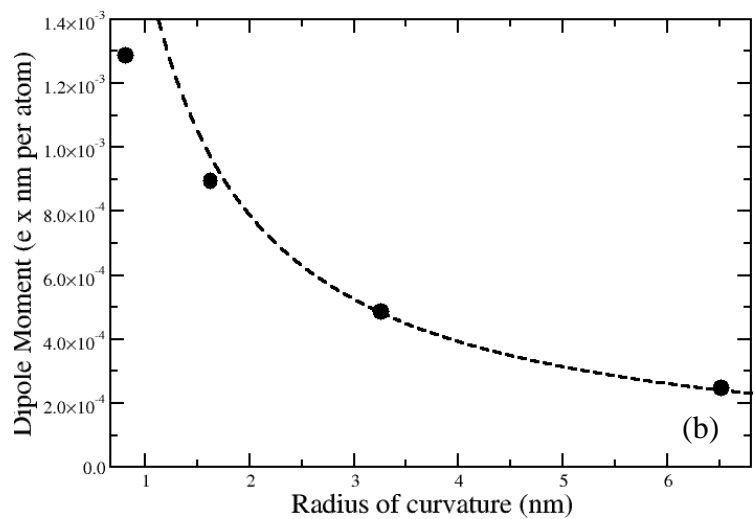
(b)
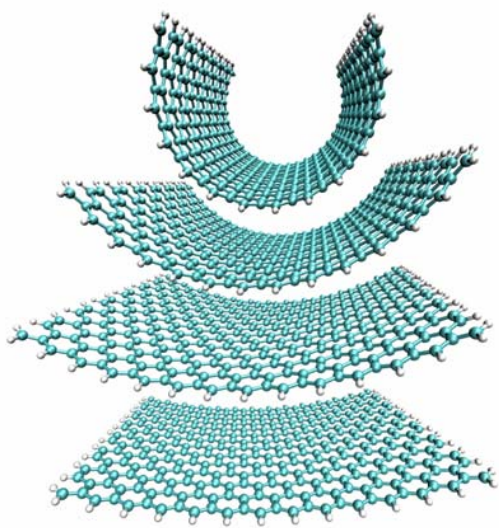
(c)
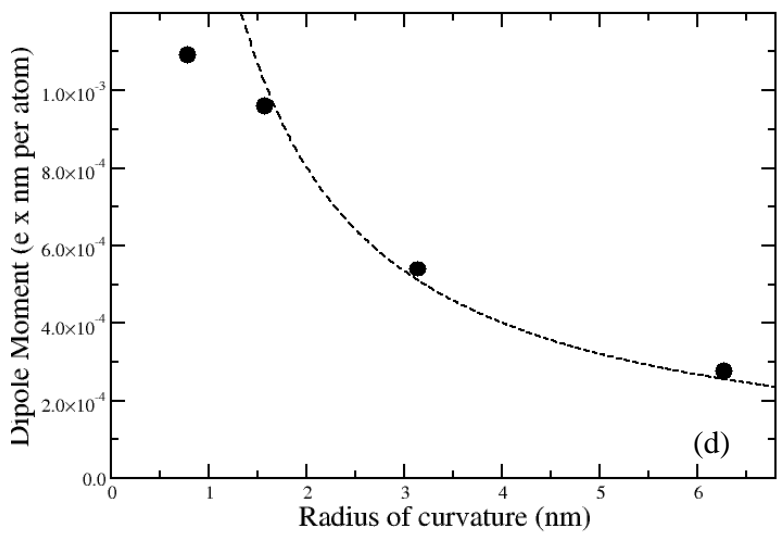
(d)

**Figure 2.** S.V. Kalinin and V. Meunier

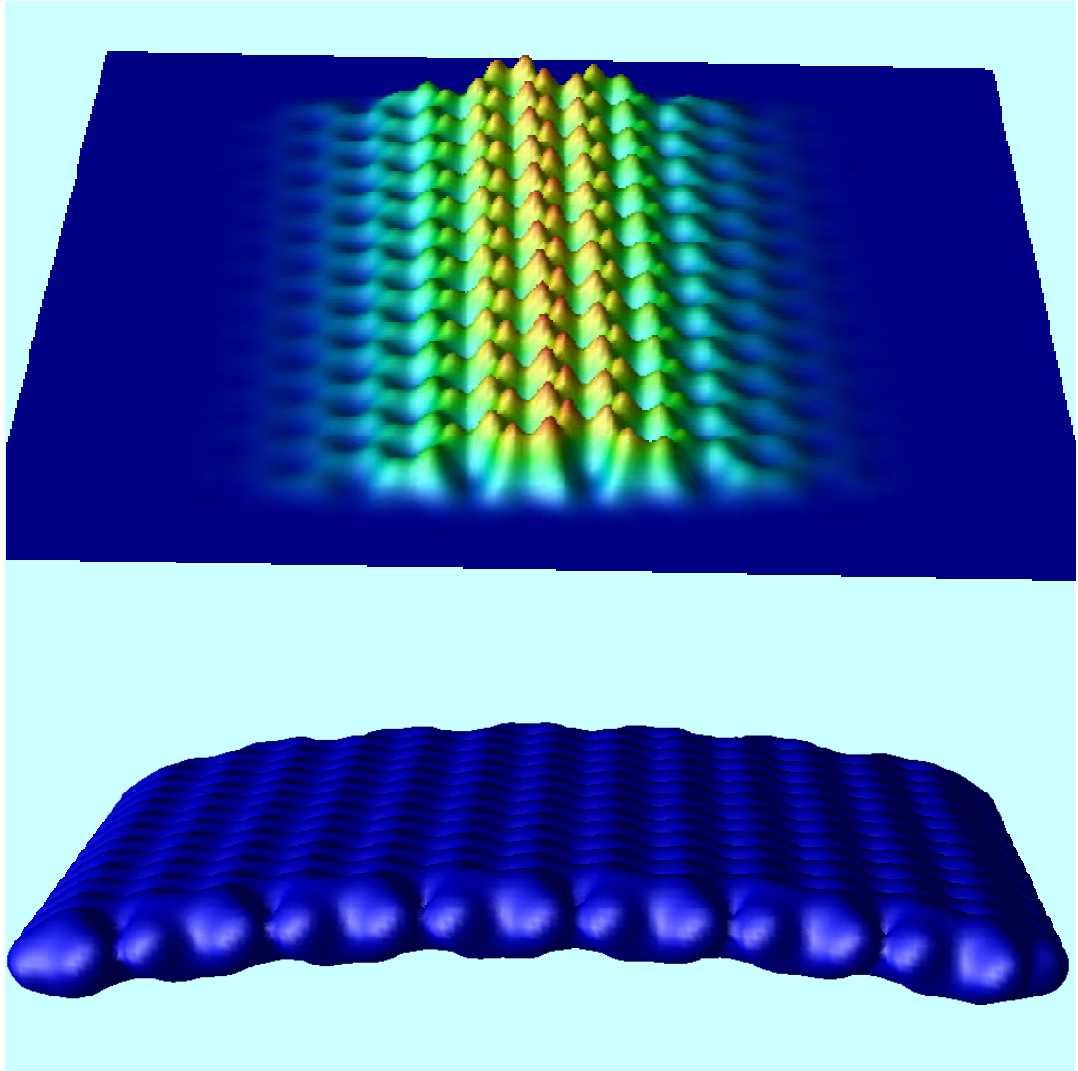

**Figure 3.** S.V. Kalinin and V. Meunier